\documentclass[conference]{IEEEtran}
\IEEEoverridecommandlockouts
\usepackage{cite}
\usepackage{amsmath,amssymb,amsfonts}
\usepackage{algorithmic}
\usepackage{graphicx}
\usepackage{textcomp}
\usepackage{xcolor}
\def\BibTeX{{\rm B\kern-.05em{\sc i\kern-.025em b}\kern-.08em
    T\kern-.1667em\lower.7ex\hbox{E}\kern-.125emX}}
\begin{document}

\title{Recognition of Tactile-related EEG Signals Generated by Self-touch\\
%\title{Recognition of Tactile-related EEG Signals Generated by Self-touch of Bare-skin\\
%\title{Conference Paper Title*\\
%{\footnotesize \textsuperscript{*}Note: Sub-titles are not captured in Xplore and
%should not be used}
\thanks{Research was funded by Institute of Information \& Communications Technology Planning \& Evaluation (IITP) grant funded by the Korea government (No. 2017-0-00451, Development of BCI based Brain and Cognitive Computing Technology for Recognizing User’s Intentions using Deep Learning; No. 2019-0-00079, Artificial Intelligence Graduate School Program, Korea University).}
}

%\author{\IEEEauthorblockN{Myoung-Ki Kim}
%\IEEEauthorblockA{\textit{Dept. Artificial Intelligence} \\
%\textit{Korea University}\\
%Seoul, Republic of Korea \\
%kim\_mk@korea.ac.kr}
%\and
%\IEEEauthorblockN{Jeong-Hyun Cho}
%\IEEEauthorblockA{\textit{Dept. Brain and Cognitive Engineering} \\
%\textit{Korea University}\\
%Seoul, Republic of Korea \\
%jh\_cho@korea.ac.kr}
%\and
%\IEEEauthorblockN{Hye-Bin Shin}
%\IEEEauthorblockA{\textit{Dept. Brain and Cognitive Engineering} \\
%\textit{Korea University}\\
%Seoul, Republic of Korea \\
%hb_shin@korea.ac.kr}
%\and
%\IEEEauthorblockN{Seong-Whan Lee}
%\IEEEauthorblockA{\textit{Dept. Artificial Intelligence} \\
%\textit{Korea University}\\
%Seoul, Republic of Korea \\
%sw.lee@korea.ac.kr}
%\and
%\IEEEauthorblockN{5\textsuperscript{th} Given Name Surname}
%\IEEEauthorblockA{\textit{dept. name of organization (of Aff.)} \\
%\textit{name of organization (of Aff.)}\\
%City, Country \\
%email address or ORCID}
%\and
%\IEEEauthorblockN{6\textsuperscript{th} Given Name Surname}
%\IEEEauthorblockA{\textit{dept. name of organization (of Aff.)} \\
%\textit{name of organization (of Aff.)}\\
%City, Country \\
%email address or ORCID}
%}

\author{\IEEEauthorblockN{Myoung-Ki Kim}
\IEEEauthorblockA{\textit{Dept. Artificial Intelligence} \\
\textit{Korea University} \\
Seoul, Republic of Korea \\
\textit{LG Display}\\
kim\_mk@korea.ac.kr} \\

\and

\IEEEauthorblockN{Jeong-Hyun Cho}
\IEEEauthorblockA{\textit{Dept. Brain and Cognitive Engineering} \\
\textit{Korea University} \\
Seoul, Republic of Korea \\
jh\_cho@korea.ac.kr} \\

\and

\IEEEauthorblockN{Hye-Bin Shin}
\IEEEauthorblockA{\textit{Dept. Brain and Cognitive Engineering} \\
\textit{Korea University} \\
Seoul, Republic of Korea \\
hb\_shin@korea.ac.kr} \\
\textit{}
}

%\author{\IEEEauthorblockN{Myoung-Ki Kim}
%\IEEEauthorblockA{\textit{Dept. Artificial Intelligence} \\
%\textit{Korea University} \\
%Seoul, Republic of Korea \\
%\textit{LG Display}\\
%kim\_mk@korea.ac.kr} \\
%
%\IEEEauthorblockN{Hye-Bin Shin}
%\IEEEauthorblockA{\textit{Dept. Brain and Cognitive Engineering} \\
%\textit{Korea University} \\
%Seoul, Republic of Korea \\
%hb\_shin@korea.ac.kr} \\
%
%\and
%
%\IEEEauthorblockN{Jeong-Hyun Cho}
%\IEEEauthorblockA{\textit{Dept. Brain and Cognitive Engineering} \\
%\textit{Korea University} \\
%Seoul, Republic of Korea \\
%jh\_cho@korea.ac.kr} \\
%\textit{}\\
%
%\IEEEauthorblockN{Seong-Whan Lee}
%\IEEEauthorblockA{\textit{Dept. Artificial Intelligence} \\
%\textit{Korea University} \\
%Seoul, Republic of Korea \\
%sw.lee@korea.ac.kr}
%}

\maketitle

\begin{abstract}
Touch is the first sense among human senses. Not only that, but it is also one of the most important senses that are indispensable. However, compared to sight and hearing, it is often neglected. In particular, since humans use the tactile sense of the skin to recognize and manipulate objects, without tactile sensation, it is very difficult to recognize or skillfully manipulate objects. In addition, the importance and interest of haptic technology related to touch are increasing with the development of technologies such as VR and AR in recent years. So far, the focus is only on haptic technology based on mechanical devices. Especially, there are not many studies on tactile sensation in the field of brain-computer interface based on EEG. There have been some studies that measured the surface roughness of artificial structures in relation to EEG-based tactile sensation. However, most studies have used passive contact methods in which the object moves, while the human subject remains still. Additionally, there have been no EEG-based tactile studies of active skin touch. In reality, we directly move our hands to feel the sense of touch. Therefore, as a preliminary study for our future research, we collected EEG signals for tactile sensation upon skin touch based on active touch and compared and analyzed differences in brain changes during touch and movement tasks. Through time-frequency analysis and statistical analysis, significant differences in power changes in alpha, beta, gamma, and high-gamma regions were observed. In addition, major spatial differences were observed in the sensory-motor region of the brain.
\end{abstract}


%\cite{xu2019phase, lee2015subject, lee2020continuous}

\section{Introduction}
Brain-computer interface (BCI) is a technology that provides means of communication and enables device control by reflecting the user's intentions through communication with various computers and machines based on the user's brain signals\cite{Jeong2020TNSRE}. Non-invasive BCI systems can also be used for diagnosis\cite{Zhang2017SciRep, Jeong2020PR}. Electroencephalography (EEG) is the most commonly used modality for recording brain signals in BCI systems because it is non-invasive and has a high temporal resolution. Non-invasive BCI systems have been studied using various types of EEG modality such as error-related potential (ErrP)\cite{MSpuler2012Clin.Neurophysiol.}, event-related potential (ERP)\cite{SKyeom2014PLoSOne, DOwon2018ITNSRE}, steady-state visual evoked potential (SSVEP)\cite{MHLee2018TNSRE, DOWon2016JNE, NSKwak2017PLoSOne}, and steady-state somatosensory evoked potential (SSSEP)\cite{KTKim2016TNSRE, LYao2017TNSRE, SSu2020JNE}, motor imagery (MI)\cite{TEKam2013Neurocomputing, OYKwon2019NNLS, TYu2015TBE}, and motor execution (ME)\cite{JHKim2015TNSRE, LWang2010Brain}.

Most BCIs are driven only by visual feedback and there is no somatosensory feedback, whereas humans have multiple sensory channels. Among the five human senses (sight, hearing, taste, smell, and touch), the sense of touch is the most extensive and the first to develop when the fetus is in the womb\cite{TField2014Touch}. The sense of touch of humans has gained relatively less attention than sight and hearing. However, with the recent development of technologies such as VR and AR, the importance and interest in haptic technology related to touch are increasing. In particular, regarding EEG-based tactile sensation, there have been several studies\cite{Greco2019JBH, APisoni2018NeuroImage, MKKim2021Winter, JHCho2021Winter, AMoungou2016SciRep, Chen2019TNSRE} that measured the surface roughness of artificial structures. However, most studies have used passive contact methods in which the objects move, while the human subject remains still. To date, there has been no EEG-based BCI research on active skin touch, whereas, in reality, humans mostly employ active touch for tactile exploration.

In this study, we collected EEG signals for tactile-related brain changes upon skin contact based on active touch. The brain signal changes according to the movement and the touch were compared and analyzed. Significant differences were observed through time-frequency analysis and statistical analysis. To the best of our knowledge, this is the first study to investigate the changes in brain dynamics upon active skin touch, which will serve as an important basis for our future research direction.

The rest of this document is organized as follows:
Section II provides a description of the experimental protocol,
EEG signal acquisition, and data analysis.
Section III provides the results of the time-frequency analysis and statistical analysis and a discussion of our work.
In session IV, conclusion are described.\\

\section{Materials and Methods}

\subsection{Participants}

Six participants (4 males; 2 females; all right-handed; age 26±6 years) were recruited for this study.
The participants had no history of neurological, psychiatric, or tactile problems.
This study was approved by the Institutional Review Board at Korea University (KUIRB-2020-0013-04), and written informed consent was obtained from all participants before the experiments. The experiments were conducted in accordance with the Declaration of Helsinki.

\subsection{Experimental Setup}

During the experimental protocol session, each participant was asked to sit on a comfortable chair in front of the experimental table.
A screen for cues and fixation was installed on the table. In order to minimize the movement of the participants' eyes and upper bodies, we asked them to fix their gaze on the fixation of the screen during the experiments.
Black barriers were installed on the front and left and right sides of the participant. This barrier is intended to prevent participants from receiving external visual stimuli during the experiments.

 The experimental procedure was performed according to a random crossover design. Participants placed their hands on the table and maintained a comfortable position.
 The experiments consisted of touch tasks and movement tasks. In the touch task, participants were asked to gently touch their left arm with their right hand. In the movement task, participants were asked to perform the same actions as in the touch task, without actually touching them.
 In both the touch task and the movement task, hand movements except for actual touch were asked to maintain the same motion.
 Participants were given sufficient time to rest between each task to minimize fatigue.

\subsection{Data Acquisition}

EEG data were recorded using a BrainAmp (BrainProduct GmbH, Germany) amplifier.
It was collected at a sampling rate of 2,500 Hz using 64 Ag/AgCl electrodes following the 10-20 international system.
At the same time, a 60Hz notch filter was used to eliminate power frequency interference.
FCz was used as the reference electrode and FPz was used as the ground electrode.
The impedance of all electrodes was maintained below 10 k$\Omega$.

\subsection{Data Analysis}

\begin{figure}[t]
\centerline{\includegraphics[width = \columnwidth]{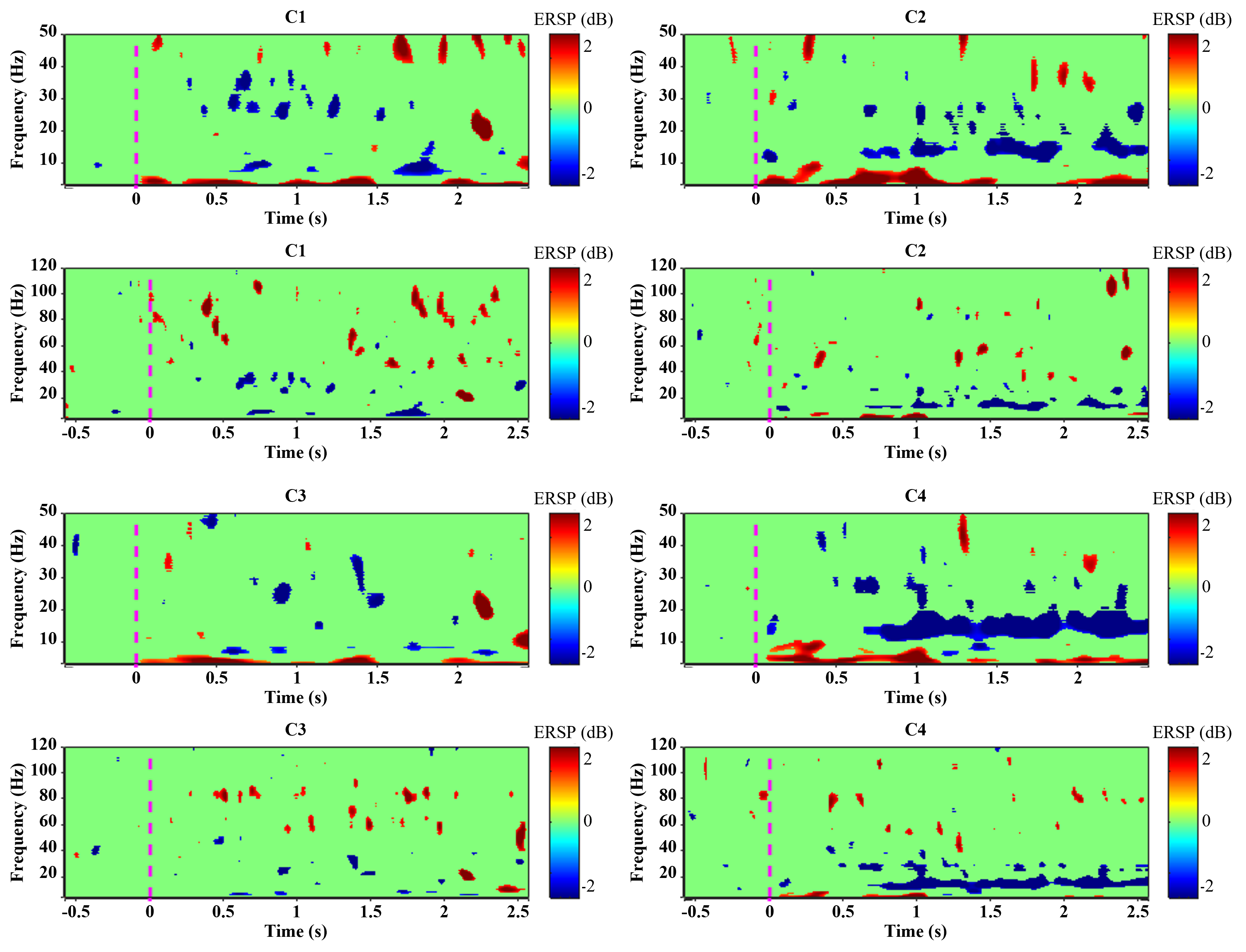}}
\caption{ERSP for channels C1, C2, C3, and C4 during the touch tasks of subject 4.}
%The baseline for calculating the ERSP was taken from the last 500 ms of the rest-phase before the touch tasks. Blue %indicates a decrease in power relative to baseline, and red indicates an increase in power versus baseline. Changes of %less than 1 \% from baseline are shown in green.
%The first row is the ERSP for channels C1 and C2 in the frequency range 0.5-50 Hz, and the second row is the ERSP for %channels C1 and C2 in the frequency range 4-120 Hz.
%The third row is the ERSP for channels C3 and C4 in the frequency range 0.5-50 Hz, and the fourth row is the ERSP for %channels C3 and C4 in the frequency range 4 to 120 Hz.
%A decrease in power occurs in the alpha and beta bands, which tends to be more pronounced in the C2 and C4 channels. An %increase in power occurs in the gamma and high-gamma bands.}
\label{fig}
\end{figure}

\begin{figure}[t]
\centerline{\includegraphics[width = \columnwidth]{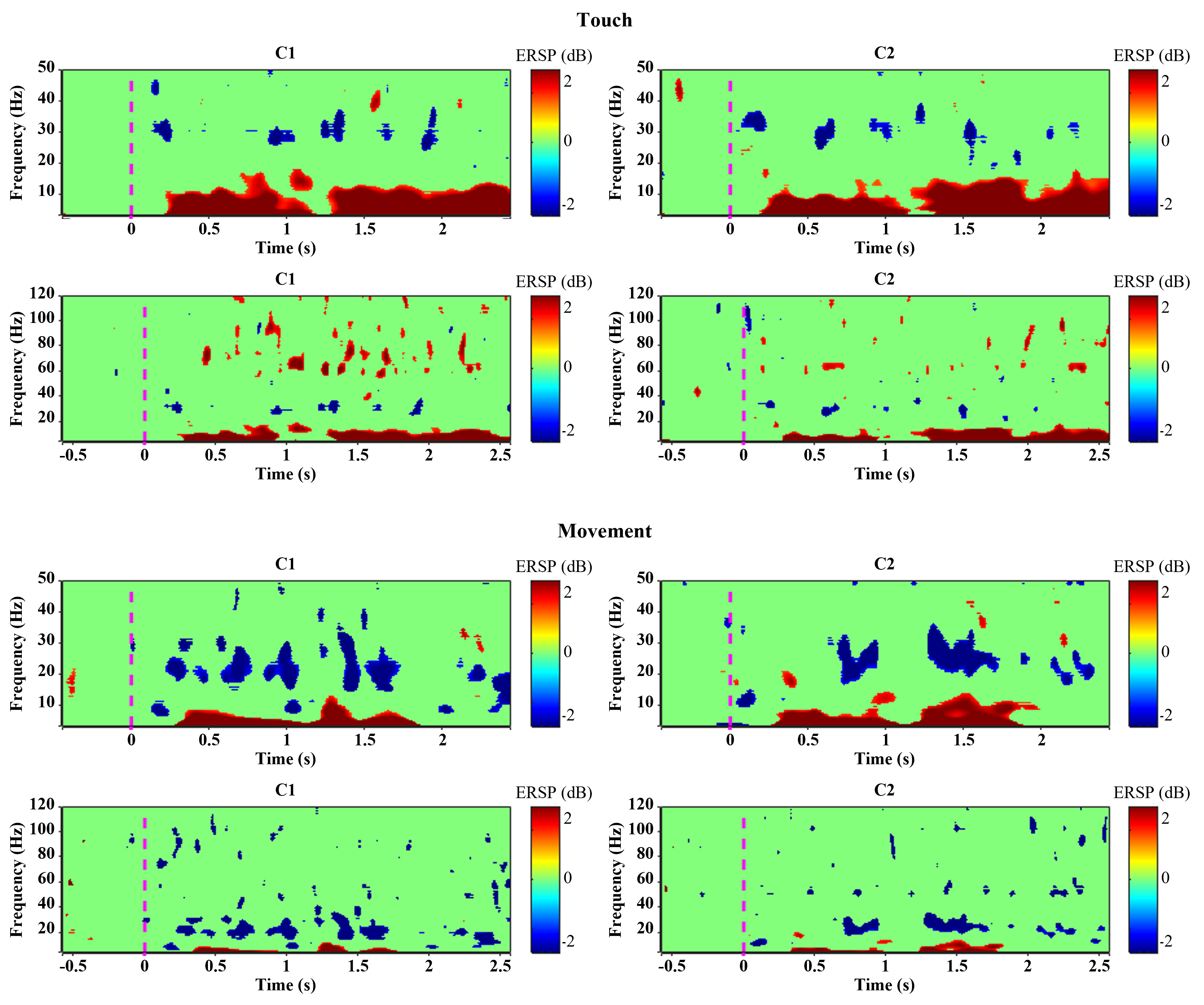}}
\caption{ERSP for channels C1 and C2 channels during the touch tasks and the movement tasks of subject 6.}
%The baseline for calculating the ERSP was taken from the last 500 ms of the rest-phase before the touch tasks and %movement tasks. Blue indicates a decrease in power relative to baseline, and red indicates an increase in power versus %baseline. Changes of less than 1 \% from baseline are shown in green.
%The first row and the second row are the ERSPs for channels C1 and C2 in the frequency range 0.5-50 Hz and 4-120 Hz, %respectively, during the touch tasks.
%The third row and the fourth row are the ERSPs for channels C1 and C2 in the frequency range 0.5-50 Hz and 4-120 Hz, %respectively, during the movement tasks.
%In the ERSP for the touch tasks, a decrease in power occurs in the (alpha and) beta bands and an increase in power %occurs in the gamma and high-gamma bands.
%In the ERSP for the movement tasks, a decrease in power occurs in the alpha and beta bands, and unlike the touch tasks, %there is no increase in power in the gamma and high-gamma bands.}
\label{fig}
\end{figure}

\begin{figure*}[t]
\centerline{\includegraphics[width = \textwidth]{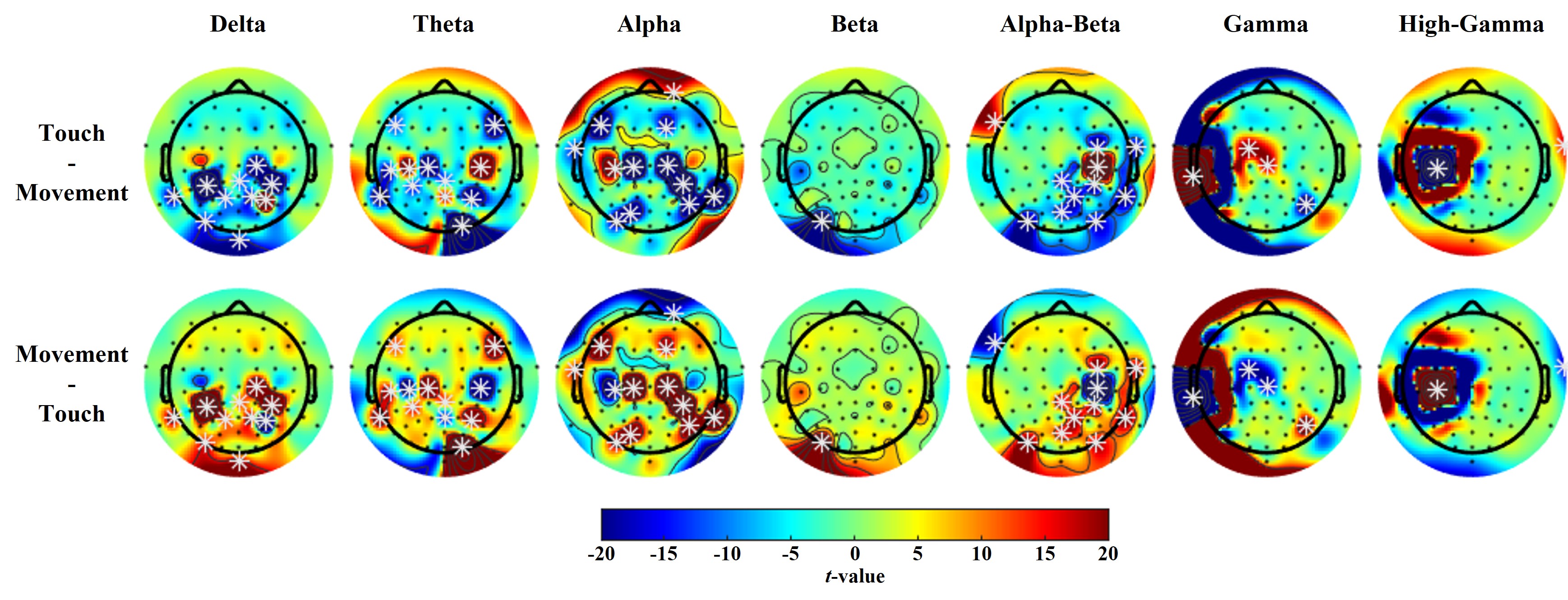}}
\caption{Differences in spectral power between touch and movement tasks. The upper line shows the spectral power differences between the touch and movement tasks when the touch task is taken as a baseline, and the lower row shows the spectral power differences between the movement and the touch tasks when the movement task is taken as a baseline. The statistical results represent t-values in each frequency band for differences between touch and movement tasks using a paired-samples t-test. The white asterisk indicates a significant electrode in spectral power between the touch and movement tasks ($p<0.05$ with Bonferroni’s correction).}
\label{fig}
\end{figure*}

The EEG data were analyzed using MATLAB R2020b with BBCI toolbox and EEGLAB toolbox (version 2021.1).
The EEG signals were downsampled to 500 Hz and bandpass filtered from 0.5 to 120 Hz.
Independent component analysis was performed to remove noise including ocular and muscle artifacts.

We conducted a comparative analysis after dividing the EEG signals collected during (i) touch task and (ii) movement task to analyze changes in the brain related to tactile sensations that occur when the skin was touched.
The preprocessed EEG signals were explored in six frequency bands as follows: delta (0.5–4 Hz), theta (4–8 Hz), alpha (8–13 Hz), beta (13–30 Hz), gamma (30–50 Hz), and high-gamma ( 50–80 Hz) bands.

\subsubsection{Time-frequency Analysis}\label{AA}
In order to understand the variance of power in the alpha, beta, and gamma bands, we performed time-frequency analysis for each channel of the data collected from the touch tasks and movement tasks.
We analyzed the change in spectral power with event-related changes at each time-frequency during each task using the event-related spectral perturbation (ERSP) method\cite{ADelorme2004JNM}.
To investigate the brain regions reported in tactile and motor related studies\cite{BMAdhikari2014NeuroImage}, we selected 22 channels (AF3, AF4, F3, F1, Fz, F2, F4, C3, C1, Cz, C2, C4, CP3, CP1, CPz, CP2, CP4, P3, P1, Pz, P2, and P4).
The baseline for calculating the ERSP was taken from the last 500 ms of the rest-phase before the touch tasks and the movement tasks. The data between 0 to 3,000 ms based on the onset point were used for the analysis.
Frequency ranges were calculated from 0.5 to 50 Hz and from 4 to 120 Hz.

\subsubsection{Spectral Power Analysis}\label{AA}
We performed the power spectral density (PSD) analysis to analyze the spectral power of each channel for the touch task and the movement task\cite{JSKumar2012PE, BMAdhikari2014NeuroImage}.
We extracted seven frequency bands using the fast Fourier transform that transforms from time to frequency domain.
The seven frequencies are as followed: delta, theta, alpha, beta, alpha-beta, gamma, and high-gamma bands\cite{EBasar2001IJP, BJRoach2008SB}.
Spectral power was calculated for each trial and then averaged across trials for each participant\cite{MLee2020Access}.

\subsubsection{Statistical Analysis}\label{AA}
A one-way analysis of variance was applied to spectral powers to investigate the comparative differences between touch tasks, movement tasks, and baseline (rest).
Paired t-test with Bonferroni's correction was used for post-hoc analysis.
All significance levels in this study were $\alpha = 0.05$.\\

\section{Results and discussion}

\subsection{Neurophysiology Analysis}\label{AA}
We collected tactile-related brain changes as EEG signals upon skin touch based on active touch. In addition, in order to distinguish the difference between movement and tactile sensation caused by active touch, the active touch task and the movement task were comparatively analyzed.
We focused on the time-frequency differences for each channel during touch and movement tasks (Fig. 1, Fig. 2).
Specifically, significant changes in power were observed in the selected 22 channels. Among them, channels C1, C2, C3, and C4 channels showed remarkable differences in power change according to the frequency domain related to tactile sensation and movement.

The ERSP results during the touch tasks mainly resulted in power reduction in the alpha and beta bands in the C1, C2, C3, and C4 channels. In all subjects, a decrease in power in the beta band was observed, but in some cases, a decrease in power in the alpha band was not observed. Although there was a difference in the degree of activation depending on the subject and task, the degree of activation mainly at C2 and C4 tended to be more prominent. Also, an increase in power occurred in the gamma and high-gamma bands. This tended to be more pronounced, mainly in the degree of activation at C1 and C3 (Fig. 1).

The ERSP results during the motor tasks were similar to those of the touch tasks: reduction of power in the alpha and beta bands, mainly in the C1, C2, C3, and C4 channels.
A decrease in power in the alpha and beta bands was observed in all subjects.
However, no increase in power occurred in the gamma and high-gamma bands (Fig. 2). Changes in gamma and high-gamma bands were identified as the biggest differences between the touch tasks and the movement tasks.

\subsection{Statistical Analysis}\label{AA}
We compared the spatial-spectral power difference at 7 frequency bands through the PSD analysis for each channel for the touch tasks and the movement tasks(Fig. 3). Statistical differences were identified in 7 frequency bands. Compared with the previously analyzed ERSP results, statistical differences were confirmed in alpha, alpha-beta, gamma, and high-gamma in the significant areas.
Significant differences in spectral power were observed mainly in the central and parietal regions, which are known to be related to touch and movement.
%In the results for the alpha region, significant differences were confirmed in the C1, C2, C3, and C4 channel regions, and it can be seen that the decrease in power in the C3 channel is larger in the movement tasks compared to the touch tasks. This can be seen as a result of the large activation of the C3 channel, the contralateral motor area, by the movement of the right hand. In the sensory-motor area except for C3, it can be seen that the power reduction of the touch tasks is larger.

In the results for the alpha-beta area, we can see that the right sensory-motor area of the brain shows a major statistical difference. In particular, it can be seen that the power change in the C4 channel is large. It can be seen that the PSD value of the touch tasks is larger than the movement task in the C4 channel.

In the results for the high-gamma regions, it can be seen that the difference in power change in the C3 region is large. This is consistent with the results of ERSP, and it can be seen as the effect of power activation in the contralateral motor area by the movement of the right hand.

\section{Conclusion}
We collected EEG signals for brain changes related to tactile sensation upon skin touch based on active touch. In addition, changes in the brain caused by movement according to active touch and changes in the brain according to tactile sensation were compared and analyzed. Significant differences were observed by performing time-frequency analysis and statistical analysis. The spatial power change and difference were observed in the sensory-motor area. Also, the difference in power change was observed in the alpha, beta, gamma, and high-gamma regions. To the best of our knowledge, These results of our research are the first results of skin touch studies based on active touch, and this will serve as a major basis for our future research direction.

\section*{Acknowledgment}

The authors thank to G.-H. Shin, B.-H. Kwon and H.-J. Ahn for help with the useful discussions of the experiment.

%\section*{References}
\bibliographystyle{IEEEbib}
\bibliography{refs}

\end{document}